\documentclass{pasj00}
\draft

\begin{document}
\SetRunningHead{J. Borovi\v cka et al.}
{Photographic observations of the Hayabusa re-entry}
\Received{2011/05/26}%{yyyy/mm/dd}
\Accepted{2011/07/19}%{yyyy/mm/dd}

\title{Photographic and Radiometric Observations of the Hayabusa Re-entry}

\author{Ji\v r\'\i\  \textsc{Borovi\v cka}}
\affil{Astronomical Institute of the Academy of Sciences, Fri\v cova 298, CZ-25165 Ond\v rejov
Observatory, \\ Czech Republic}
\email{Jiri.Borovicka@asu.cas.cz} 

\author{Shinsuke \textsc{Abe}}
\affil{Institute of Astronomy, National Central University, 300 Jhongda Road, Jhongli, Taoyuan, 32001, Taiwan }
\email{shinsuke.avell@gmail.com}

\author{Luk\'a\v s  \textsc{Shrben\'y}}
\affil{Astronomical Institute of the Academy of Sciences, Fri\v cova 298, CZ-25165 Ond\v rejov
Observatory, \\ Czech Republic}

\author{Pavel \textsc{Spurn\'y}}
\affil{Astronomical Institute of the Academy of Sciences, Fri\v cova 298, CZ-25165 Ond\v rejov
Observatory, \\ Czech Republic}

\author{ Philip A. \textsc{Bland}}
\affil{IARC, Department of Earth Science and Engineering,
Imperial College London, London SW7 2AZ, UK}


\maketitle

\begin{abstract}
We analyzed photographic observations of the re-entry of the Hayabusa spacecraft and capsule
over Southern Australia on June 13, 2010, 13:52 UT. Radiometric measurements of the brightness of the
associated fireball were obtained as well. We derived the trajectories and velocities of
the spacecraft, its four fragments and the capsule. The capsule trajectory was within a few hundred 
meters of the trajectory predicted by JAXA prior the re-entry. The spacecraft trajectory was about 
1 km higher than the capsule trajectory. Two major fragments separated from the spacecraft at a 
height of about 62 km
with mutual lateral velocity of 250 m/s. The maximum absolute magnitude of the fireball 
of $-12.6$ was reached at a height of 67 km.  The dynamic pressures acting on the spacecraft
at the fragmentation points were only 1 -- 50 kPa. No spacecraft
fragment was seen to survive below the height of 47 km. The integral luminous efficiency of the event was 1.3\%. As expected, the capsule had a very low luminous efficiency and very low 
ablation coefficient. The ablation coefficients and masses of the major spacecraft fragments
are discussed.

\end{abstract}

\section{Introduction}

We present results of photographic observation of the Hayabusa reentry from two temporary stations and two stations of the Desert Fireball Network located in Southwestern Australia. All cameras used wide field (fish-eye) lenses and captured the artificial fireball produced by the reentry of Hayabusa on single sheets of films. The Desert Network observatories are also equipped with all-sky radiometers measuring the fireball luminosity. Standard astrometric and photometric procedures could be used to obtain the trajectory, velocity and brightness of the fireball. However, most of the spacecraft fragments were not resolved. Only a few widely separated and bright enough fragments could be tracked individually.

\section{Observation and data reduction}

Two temporary stations were set up in the desert of South Australia by a ground observation team of the Japan Aerospace Exploration Agency (JAXA) \citep{Fujita} to observe the re-entry. They were located to the north and to the south of the expected trajectory, respectively. Among other instruments, the stations were equipped by photographic cameras MAMIYA $6\times7$ carrying the fish-eye lens $f=17$ mm, $f/4$ (GOS3) and $f=24$ mm $f/2.8$ (GOS4) and using Fujifilm color 400 ASA.  The northern station Coober Pedy, designed GOS4 and operated by Y. Kakinami and Y. Shiba, was equipped with a rotating shutter placed in front of the camera lens and producing 10 breaks per second for measuring the fireball velocity. The southern station Tarcoola, GOS3, was operated by S. Abe. At both stations the cameras were tilted from zenith toward the fireball trajectory (by 33$^\circ$
and 45$^\circ$, respectively).

\begin{figure}
\FigureFile(\linewidth,1mm){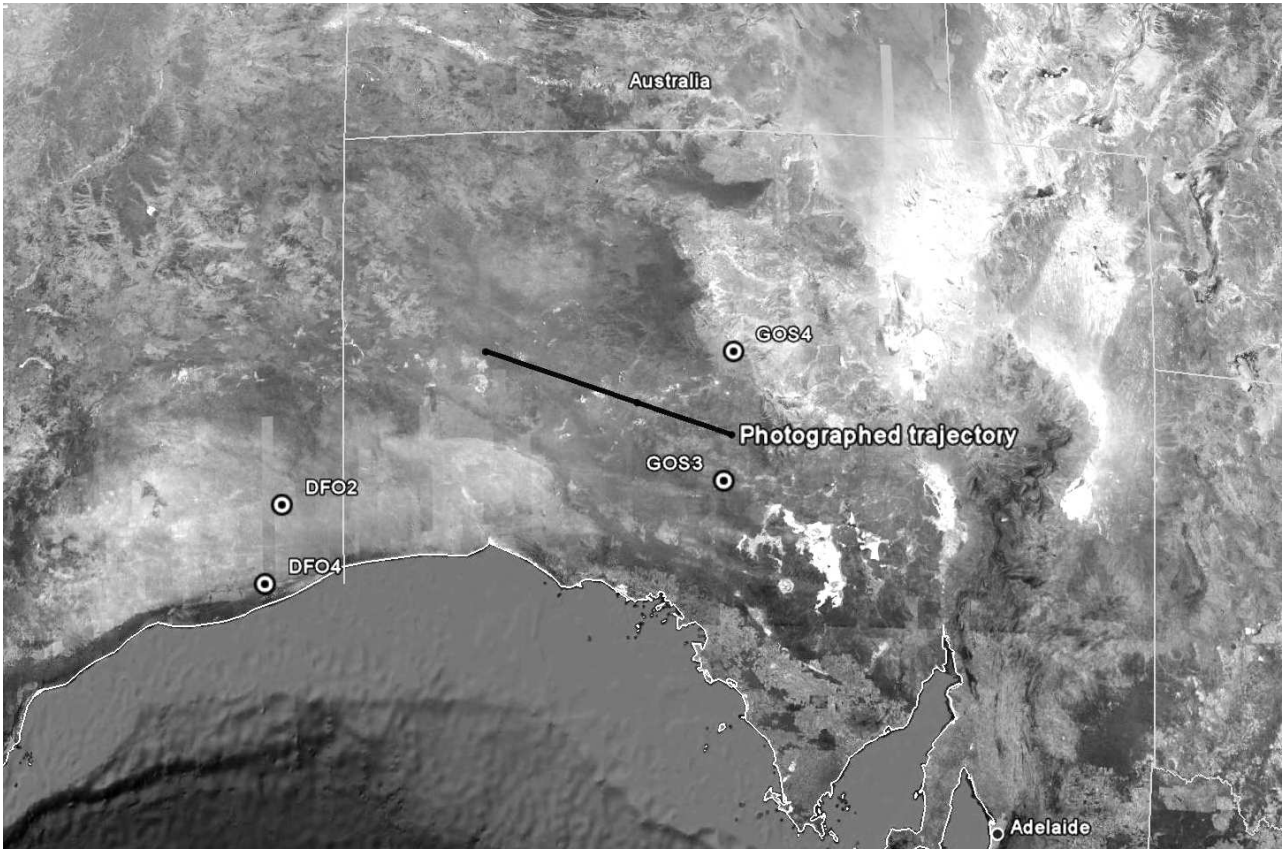}
\caption{The positions of observing stations GOS3, GOS4, DFO2, and DFO4 on the map
of southern Australia. The photographed part of Hayabusa trajectory is shown as well. The ground projection of the photographed trajectory is 365 km long.}
\label{map}
\end{figure}

\begin{figure*}
\FigureFile(\linewidth,1mm){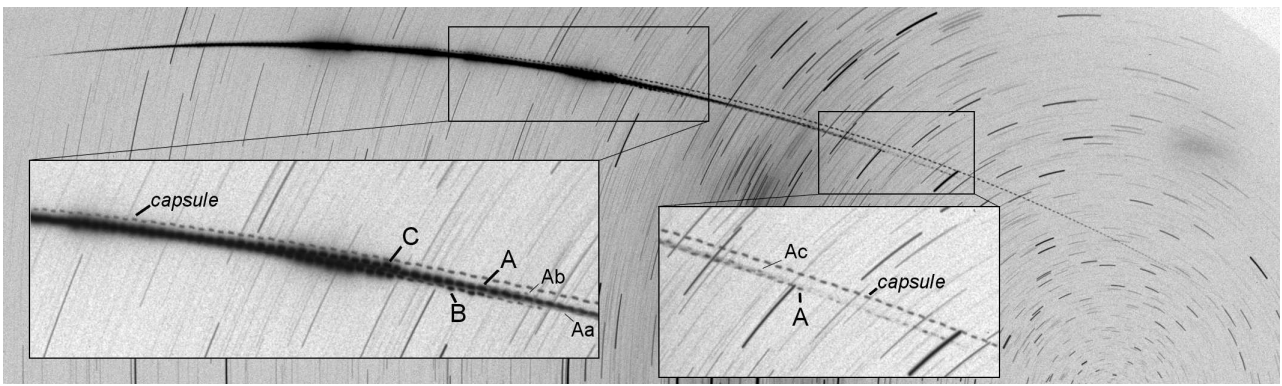}
\caption{Hayabusa re-entry as photographed from station GOS4. The exposure was 27 minutes long, from 13:37:00 to 14:04:00 UT.  The fireball flew from left to right and was interrupted be the rotating shutter 10 times per second. The closest horizon lies upwards.  The fragments mentioned 
in the text are identified in the insets. }
\label{GOS}
\end{figure*}

\begin{figure}
\begin{center}
\FigureFile(140mm,140mm){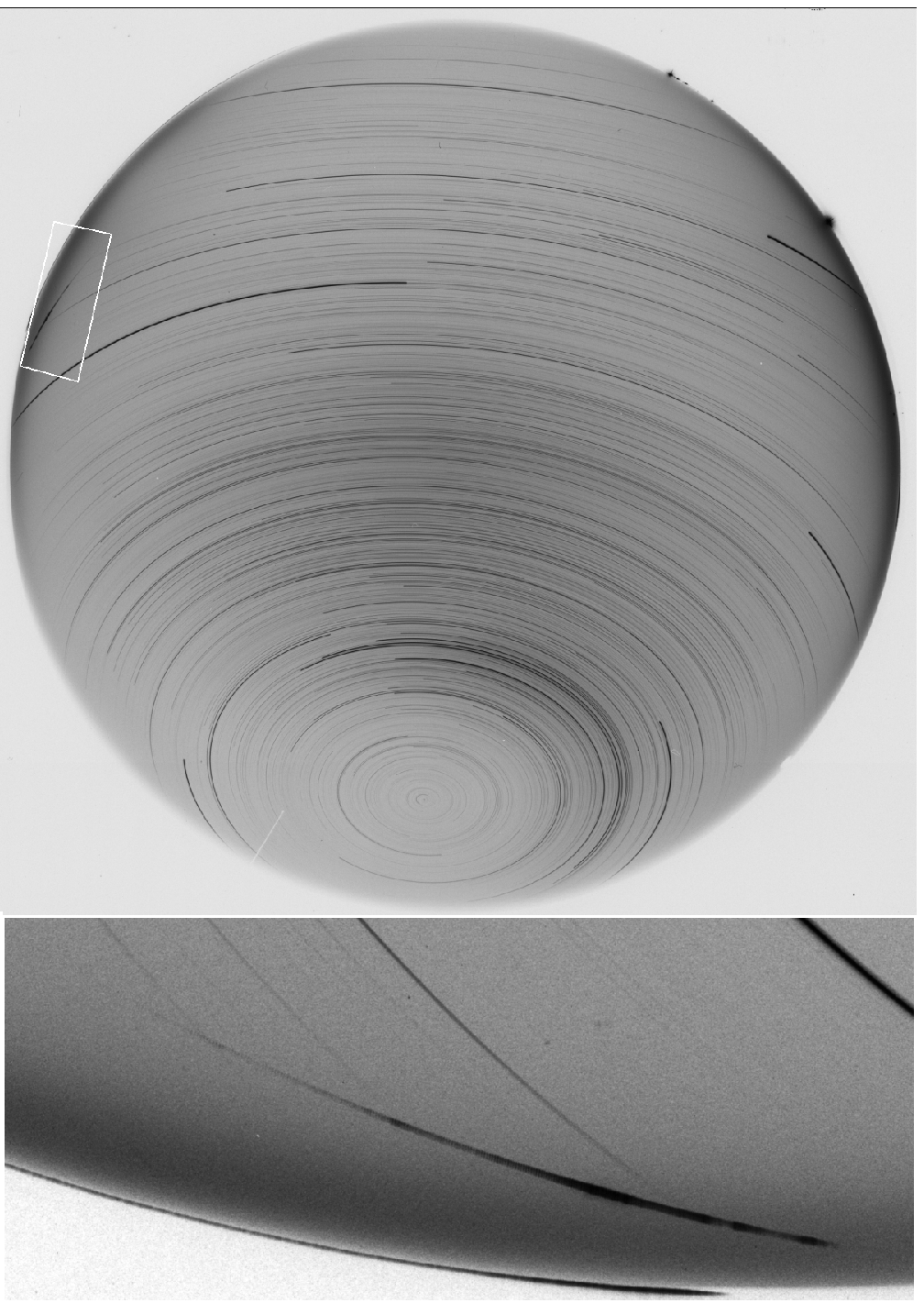}
\caption{All-sky image from DFO2 camera showing the Hayabusa re-entry fireball above the ENE horizon (top) and enlargement of the fireball (bottom). The fireball flew from left to right. The exposure was 11 hours 55 minutes long (9:30 -- 21:25 UT). Shutter breaks (15 per second) are  visible on original records along most of the image. The fireball ends only 3.2$^\circ$ above horizon. }
\label{DFO}
\end{center}
\end{figure}

The Desert Fireball Network (DFN) is located in the Nullarbor Plain in Western Australia and consists of four Autonomous Fireball Observatories \citep{Bland}. The purpose of the network is to observe natural fireballs and possible meteorite falls. The observatories are equipped with Zeiss Distagon 3.5/30 mm fish-eye lenses, producing 8-cm-diameter all-sky images on Ilford FP 125 black-and-white films. A rotating shutter is located inside of the cameras and produces 15 breaks per second. The radiometers based on photomultipliers measure the total brightness of the sky 500 times per second.

Despite the relatively large range to the Hayabusa trajectory (400--600 km), all four stations
of the DFN captured at least part of the fireball. In this paper, we use data from the two closest stations, DFO2, Forrest, and DFO4, Mundrabilla. The positions of the observing stations are shown in Fig.~\ref{map} and listed in Table~\ref{coord}.

\begin{table*}
\caption{Coordinates of the observing stations}
\begin{center}
\begin{tabular}{llll} \hline
Station & Longitude & Latitude & Altitude \\
& deg & deg & km \\ \hline
GOS3 & 134.55858 & $-$30.69911 & 0.152 \\
GOS4 & 134.71819 & $-$29.03392 & 0.224 \\
DFO2 & 128.11506 & $-$30.85808 & 0.161 \\
DFO4 & 127.84869 & $-$31.83564 & 0.085 \\
\hline
\end{tabular}
\label{coord}
\end{center}
\end{table*}

Figure~\ref{GOS} shows the photograph from station GOS4; Fig.~\ref{DFO} from station DFO2. The original photographs were scanned on photogrammetric scanner Vexcel Ultrascan 5000 and measured manually with our computer program Fishscan. The method of \citet{redsky}  was used for positional reduction of fish-eye images. Photometry was done by
measuring the image darkening with Fishscan in comparison with star trails (see \cite{Spurny2006} for discussion of the method).

The GOS images resolved the re-entry capsule from the spacecraft (see Fig.~\ref{GOS}). The spacecraft split into three fragments. Two of them (designed B and C) were bright but short-lived. The fragment A was the most durable. Later, other fragments (Aa, Ab, Ac) separated from the A trajectory. However, Aa and Ab were visible only on the GOS4 image. On the other hand, GOS3 showed Ac to split in three trails at a later stage but this splitting was not visible in GOS4. In summary, the trajectories of six objects observed from at least two stations (the capsule, the spacecraft before fragmentation, and fragments A, B, C, and Ac) were determined independently. The trajectory of each object was computed by the least squares method of \citet{mimo}. The method, in its ususal form, assumes the fireball trajectory to be a straight line and computes it by minimizing the distances between the trajectory and the lines of sight.  However, the straight line assumption is not valid for long nearly horizontal fireballs, which was the case of Hayabusa, in particular the spacecraft, the capsule and fragment A. Trajectory curvature caused by Earth gravity was significant. We therefore used a modification of the method assuming that the trajectory is
a circular arc lying in the vertical plane. For the capsule, even this approximation proved insufficient and we had to split the trajectory into several parts and compute them separately. All computations were performed in the inertial coordinate system with the origin at the center of  Earth. In this system, the coordinates of the observing stations varied with time due to Earth rotation.
Photographic data do not provide absolute timing because of long exposures. Only relative time from counting of shutter breaks is available. However, the radiometers provided the fireball light curve as a function of absolute time. By adjusting the photographic  and radiometric light curves,  we were able to obtain the absolute timing of shutter breaks  with the precision of about 0.1 second.  This method was, nevertheless, available only for the spacecraft. The capsule was on an independent  trajectory and no separate radiometric light curve for the capsule exists.

\begin{table*}
\caption{Selected points along the spacecraft trajectory}
\begin{center}
\begin{tabular}{rrrr} \hline
Time$^{\rm a)}$ & Longitude & Latitude & Height \\
\multicolumn{1}{c}{s} & \multicolumn{1}{c}{deg} &
\multicolumn{1}{c}{deg} & \multicolumn{1}{c}{km} \\ \hline
\multicolumn{4}{c}{\textit{Spacecraft}} \\
 beg  & 130.7684 & $-28.9127$ & 106.19 \\
$-3.4$& 131.1056 & $-29.0243$ &  99.88 \\
$-2.0$& 131.2609 & $-29.0753$ &  97.02 \\
  0.0 & 131.4783 & $-29.1462$ &  93.07 \\
  2.0 & 131.6966 & $-29.2170$ &  89.16 \\
  4.0 & 131.9153 & $-29.2874$ &  85.30 \\
  6.0 & 132.1348 & $-29.3576$ &  81.48 \\
  8.0 & 132.3552 & $-29.4276$ &  77.71 \\
 10.0 & 132.5775 & $-29.4977$ &  73.96 \\
 12.0 & 132.7984 & $-29.5668$ &  70.29 \\
 14.0 & 133.0168 & $-29.6347$ &  66.72 \\
 15.9 & 133.2220 & $-29.6981$ &  63.41 \\
 end$^{\rm b)}$  & 133.3510 & $-29.7377$ &  61.35 \\
\multicolumn{4}{c}{\textit{Fragment A}} \\
 17.2 & 133.3590 & $-29.7401$ &  61.26 \\
 18.0 & 133.4439 & $-29.7657$ &  59.95 \\
 20.0 & 133.6506 & $-29.8279$ &  56.79 \\
 22.0 & 133.8504 & $-29.8877$ &  53.78 \\
 24.0 & 134.0316 & $-29.9417$ &  51.08 \\
 26.1 & 134.1956 & $-29.9907$ &  48.65 \\
 27.7 & 134.2981 & $-30.0216$ &  47.14 \\
\multicolumn{4}{c}{\textit{Fragment B}} \\
 19.4 & 133.5903 & $-29.8095$ &  57.98 \\
 20.0 & 133.6509 & $-29.8274$ &  57.12 \\
 21.0 & 133.7461 & $-29.8555$ &  55.77 \\
 22.0 & 133.8338 & $-29.8814$ &  54.54 \\
\multicolumn{4}{c}{\textit{Fragment C}} \\
 18.8 & 133.5284 & $-29.7917$ &  58.48 \\
 19.0 & 133.5500 & $-29.7982$ &  58.12 \\
 20.0 & 133.6535 & $-29.8291$ &  56.41 \\
 20.5 & 133.7035 & $-29.8440$ &  55.58 \\
\multicolumn{4}{c}{\textit{Fragment Ac}} \\
 23.4 & 133.9776 & $-29.9252$ &  51.66 \\
 24.0 & 134.0304 & $-29.9407$ &  50.85 \\
 24.9 & 134.1022 & $-29.9618$ &  49.75 \\
 25.6 & 134.1551 & $-29.9774$ &  48.95 \\
\hline
\end{tabular}
\\ {\footnotesize  a) Seconds after 13:52:00 UT
\\  b) The last point where the spacecraft trail was not split
}
\label{res1}
\end{center}
\end{table*}

\begin{table*}
\caption{Selected points along the capsule trajectory}
\begin{center}
\begin{tabular}{rrrr} \hline
Time$^{\rm a)}$ & Longitude & Latitude & Height \\
\multicolumn{1}{c}{s} & \multicolumn{1}{c}{deg} &
\multicolumn{1}{c}{deg} & \multicolumn{1}{c}{km} \\ \hline
 beg  & 132.9673 & $-29.6209$ &  66.47 \\
 16.0 & 133.0768 & $-29.6545$ &  64.71 \\
 19.0 & 133.4016 & $-29.7532$ &  59.60 \\
 22.0 & 133.7122 & $-29.8466$ &  54.81 \\
 25.0 & 134.0014 & $-29.9328$ &  50.43 \\
 28.0 & 134.2593 & $-30.0092$ &  46.58 \\
 32.0 & 134.5429 & $-30.0921$ &  42.39 \\
 34.8 & 134.6962 & $-30.1367$ &  40.09 \\
end$^{\rm b)}$ & (134.7368) & ($-30.1486$) &  (39.47) \\
\hline
\end{tabular}
\\ {\footnotesize  a) Relative time
\\  b) Coordinates are less relaible because of uncertain timing
}
\label{res2}
\end{center}
\end{table*}

\section{Results}

\subsection{Trajectory}

Table~\ref{res1} gives the resulting coordinates of selected points measured along the trajectories of the Hayabusa spacecraft and its fragments. Table~\ref{res2} gives the analogous data for the re-entry capsule. In the latter case the timing is only relative. The estimated precision of our coordinates is 100 meters. 

\begin{figure}
\FigureFile(\linewidth,1mm){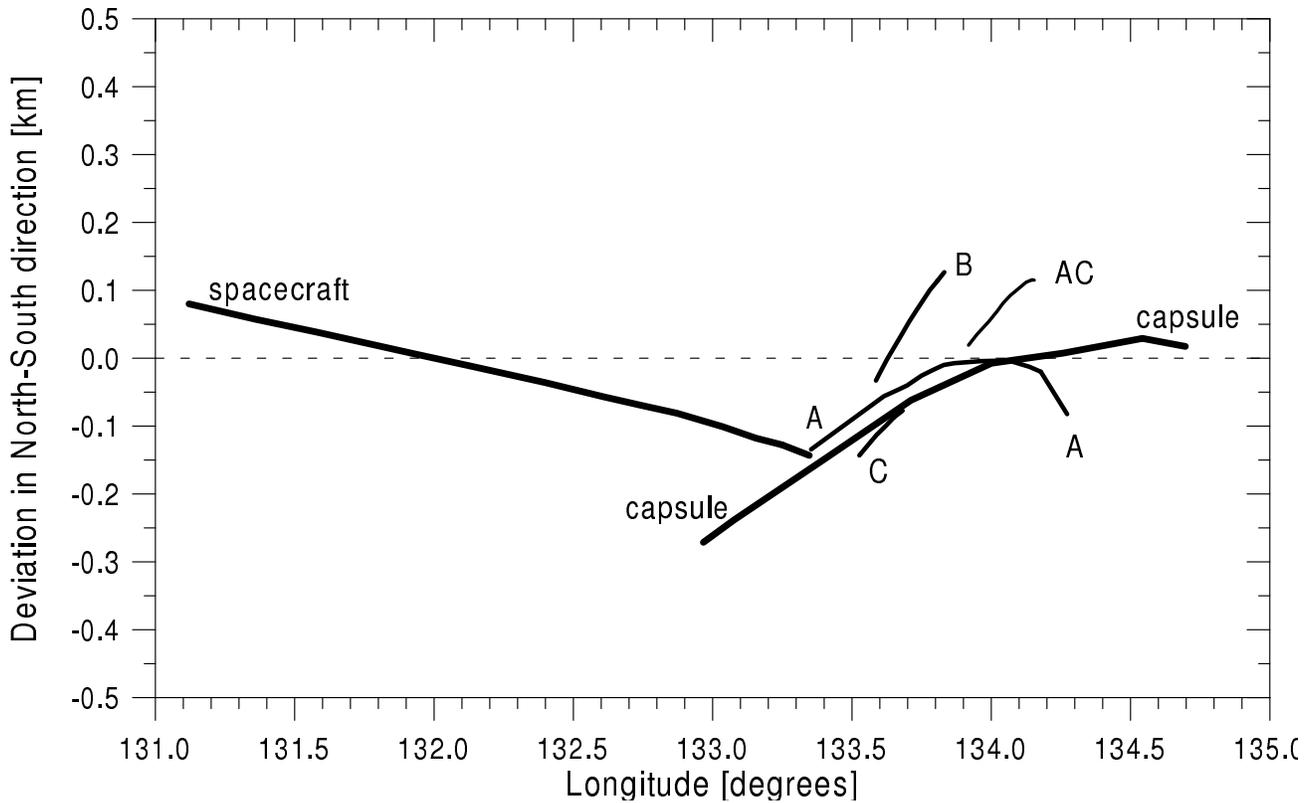}
\caption{Deviation of the observed latitude of the spacecraft, its fragments
and the capsule from the JAXA prediction for the capsule as a function of observed longitude. The error of trajectory determination is about 0.1 km. Note that fragment A was in fact a whole swarm of fragments not resolved in photographic data.}
\label{diffi}
\end{figure}

\begin{figure}
\FigureFile(\linewidth,1mm){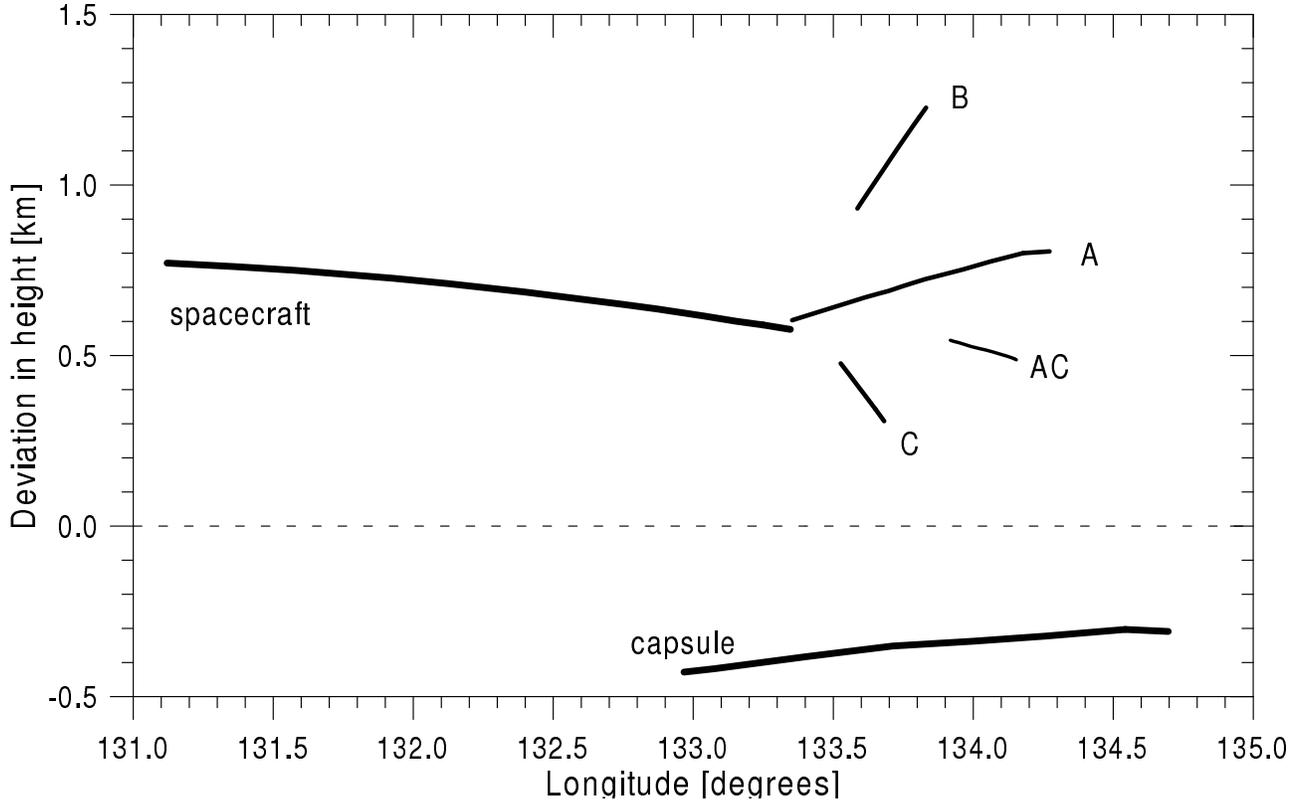}
\caption{Deviation of the observed height of the spacecraft, its fragments
and the capsule from the JAXA prediction for the capsule as a function of observed longitude. The error of trajectory determination is about 0.1 km.}
\label{difh}
\end{figure}

We compared our results with the trajectory of the capsule predicted by JAXA just before the re-entry \citep{Fujita}. 
Both our computation and the JAXA prediction were performed in the WGS 84 coordinate system.
Figure~\ref{diffi} shows the difference between
the observed and predicted latitude (expressed in km, i.e.\ as the deviation in the North-South direction) as a function of longitude. Similarly, Fig.~\ref{difh} shows the difference between the observed and predicted height above the surface. In the N-S direction, both the spacecraft, its fragments and the capsule remained within 250 meters of the predicted trajectory in the observed parts of the trajectories. After the spacecraft fragmentation, all observed fragments turned a little bit to the north. Such an asymmetry
may be due to spacecraft rotation. As for the height, the spacecraft moved about 1 km higher than the capsule and the capsule was about 300 meters lower than predicted. After the spacecraft fragmentation, fragment B turned up while fragment C turned down. This geometrical separation, together with their high brightness, made both fragments easily resolvable in our data. The divergence angle between fragments B and C was 1.2 $\pm$ 0.3
degrees, corresponding to their mutual velocity of separation of 250 
$\pm$ 60 m/s. 
The source of the lateral velocity may be a momentum gained during the explosion or differences
in the lift coefficients of various spacecraft components.
Fragments A and Ac followed more closely the original direction of the spacecraft. Video data published on the JAXA digital archive 
(Osaka Science Museum/National Central University/GOTO Inc./JAXA)\footnote{http://jda.jaxa.jp} 
show that, in fact, large number of fragments was present in this region; most of them remained unresolved in our photographs.

\begin{figure}
\FigureFile(\linewidth,1mm){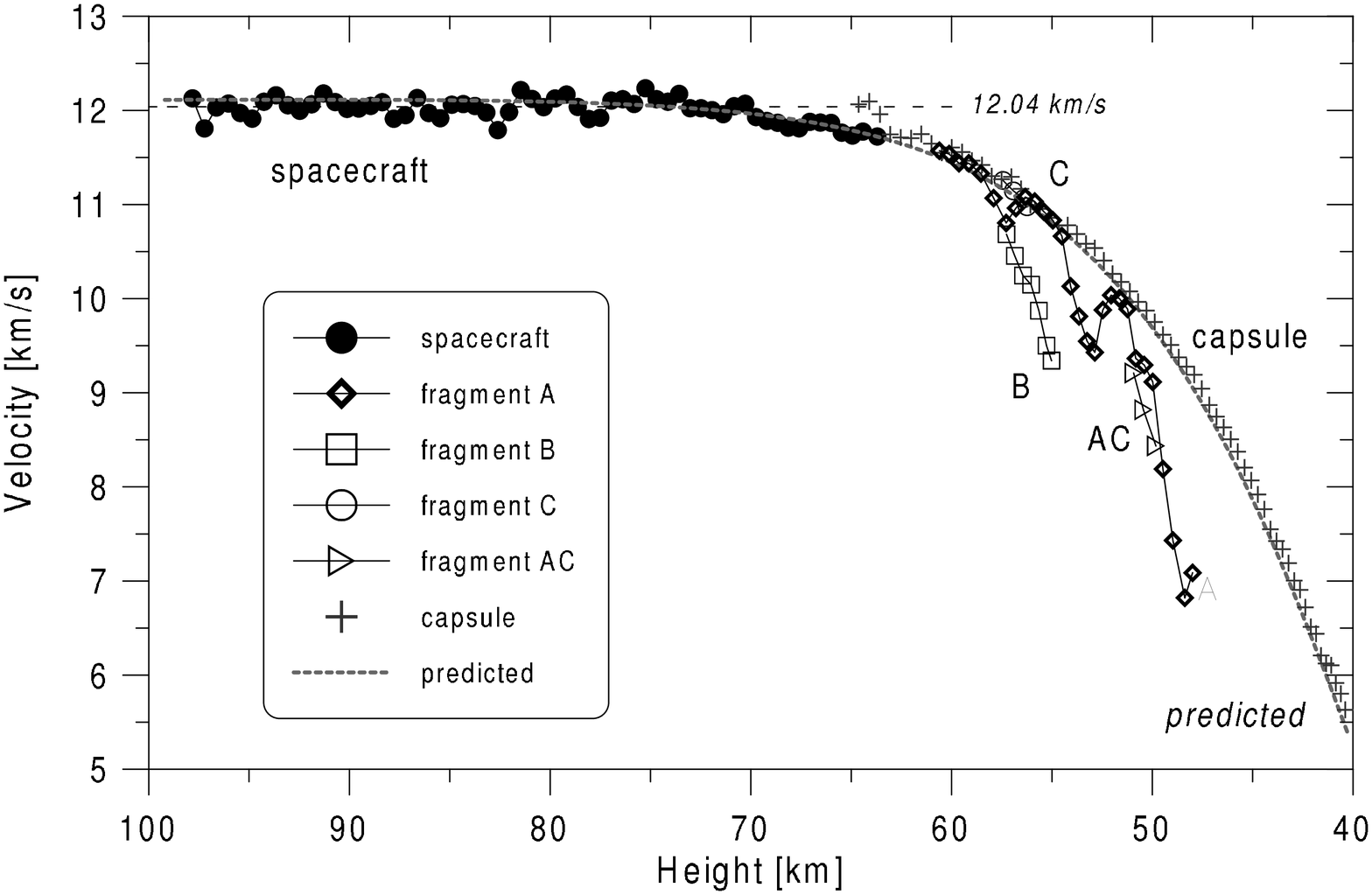}
\caption{Measured velocity (relative to the Earth center) of the spacecraft, its fragments and the capsule as a function of height above the surface compared to the JAXA prediction for the capsule. Each velocity point was computed from 9 consecutive time marks on the GOS4 image. The points are given in interval of 3 time marks, i.e. 0.3 second. The DFO2 and DFO4 images gave similar results for the spacecraft but with larger scatter.}
\label{vel}
\end{figure}

\subsection{Velocity}

Figure~\ref{vel} shows the observed velocities as a function of height. In order to smooth 
the velocity profile, running averages computed from 9 consecutive shutter breaks 
were plotted, with the step of 3 shutter breaks.
The  velocity relative to the Earth's center is given (the velocity relative to the surface was about 
0.37 km/s lower because of Earth's rotation).  The initial velocity of the spacecraft was 
12.04 km/s. This 
value was computed as the average velocity between the heights 95 and 75 km, i.e. in the interval 
where the data were good and the deceleration was negligible. Below 75 km the deceleration 
started to be noticeable.

The capsule was a much fainter object and started to be visible on the photographs at the 
height of 66.5 km, when the deceleration was already in progress. 
The velocity data at the beginning 
are rather uncertain. The first good data point is at the height of 63 km, where the velocity was 
11.75 km/s. The deceleration was then smooth and closely followed the JAXA prediction, although 
the velocity was systematically higher by about 0.1 km/s than predicted over the whole observed
arc down the height of 40 km. The last measured velocity was 5.63 km/s at the height of 40.4 km.

The last piece of fragment A ceased to be visible on our photographs at the height of 47 km; its velocity at that time was about 7 km/s. The bright fragments B and C disappeared at heights around 55 km. Fragment B decelerated much more than fragment C.

The trajectory and initial velocity of the spacecraft were used for computation of geocentric and heliocentric orbits of Hayabusa at the moment of re-entry.
The data are given in Table~\ref{orbits}.

\begin{table*}
\caption{Geocentric and heliocentric orbits of the spacecraft at the beginning of re-entry}
\begin{center}
\begin{tabular}{ll} \hline
\multicolumn{2}{l}{\textit{Geocentric orbit}} \\ \hline
Eccentricity & 1.32 \\
Inclination & $34.52^\circ$ \\
Right ascension of the ascending node & $7.58^\circ$ \\
Pericenter distance & 6310 km \\
Longitude of pericenter & $255.58^\circ$ \\
\hline
\multicolumn{2}{l}{\textit{Heliocentric orbit (J2000.0)}} \\ \hline
Semimajor axis & 1.278 AU \\
Perihelion distance &0.9824 AU \\
Eccentricity & 0.231 \\
Inclination & $1.59^\circ$ \\
Argument of perihelion & $145.63^\circ$ \\
Longitude of the ascending node & $82.360^\circ$ \\
\hline
\end{tabular}
\label{orbits}
\end{center}
\end{table*}

\subsection{Brightness}

\begin{figure}
\FigureFile(\linewidth,1mm){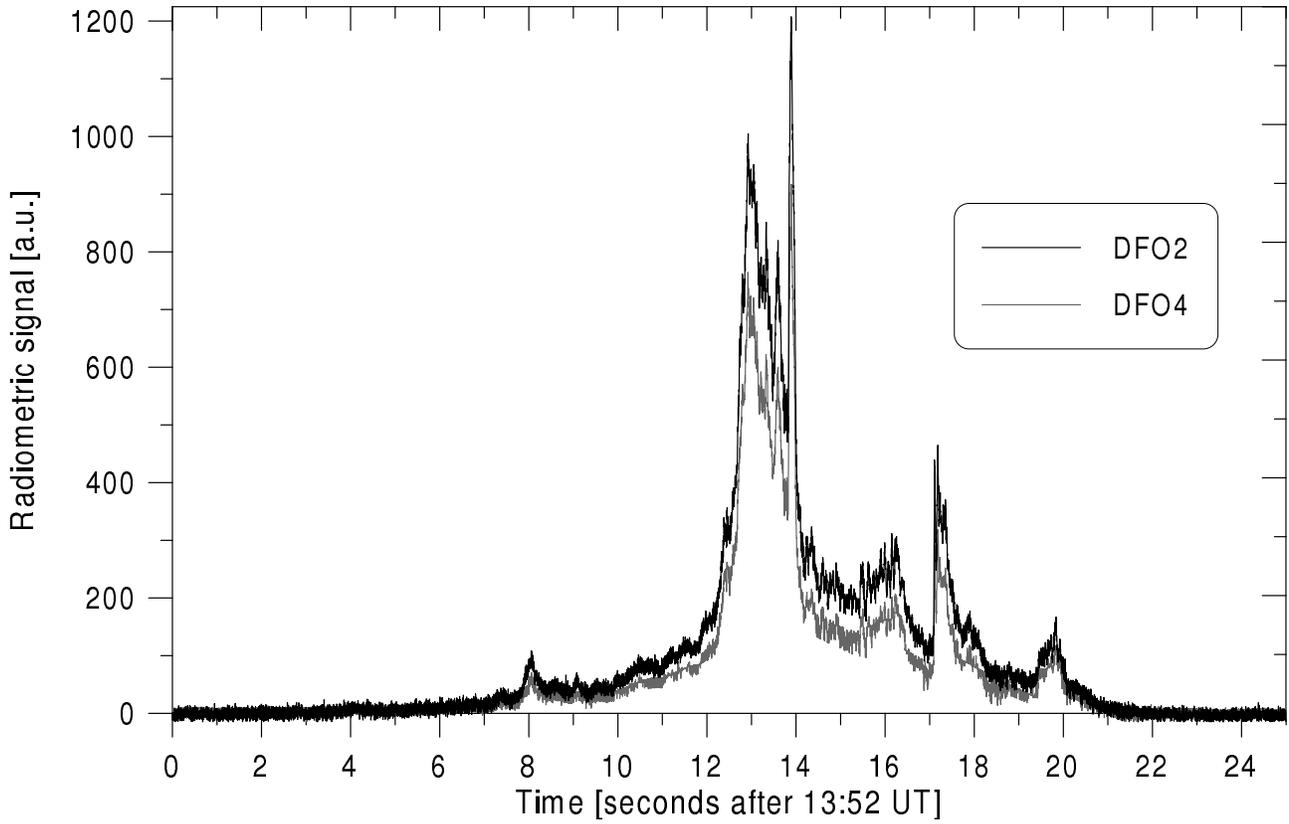}
\caption{Radiometric light curves of the Hayabusa fireball recorded on the DFO2 (upper black curve) and DFO4 (lower gray curve) stations.}
\label{lc}
\end{figure}

We performed photometric measurements at stations DFO2 and DFO4, where the individual fragments and the capsule were not resolved and total brightness of the fireball was therefore measured, and at station GOS3, where the fragments were measured individually. The fragment measurement was, however, possible only after the fragments became well separated from other fragments, i.e. not at all points where the positions and velocities were measured. More importantly, the fireball in its bright part was saturated in the GOS3 image, and the resulting brightness was underestimated. At DFO2 and DFO4, on the other hand, the calibration was a problem in the second part of the trajectory because the proximity of the fireball to the horizon. The brightness seems to be overestimated here. The photometry of the whole fireball was therefore best obtained from the radiometric data. Figure~\ref{lc} shows the radiometric light curves in linear scale  relative units as a function of time. Though there is some noise in the data, the similarity of the DFO2 and DFO4 curves demonstrates that many features seen on the light curves are real. The conversion of the radiometric signal to absolute magnitude was done using the known response of radiometers as a function of zenith distance of the object and the known range to the fireball. The scale was then adjusted using the photographic photometry at GOS3 and DFO2 for the fireball heights around 80 km, where both photographic records were without calibration problems and in good mutual agreement.

\begin{figure*}
\begin{center}
\FigureFile(\linewidth,1mm){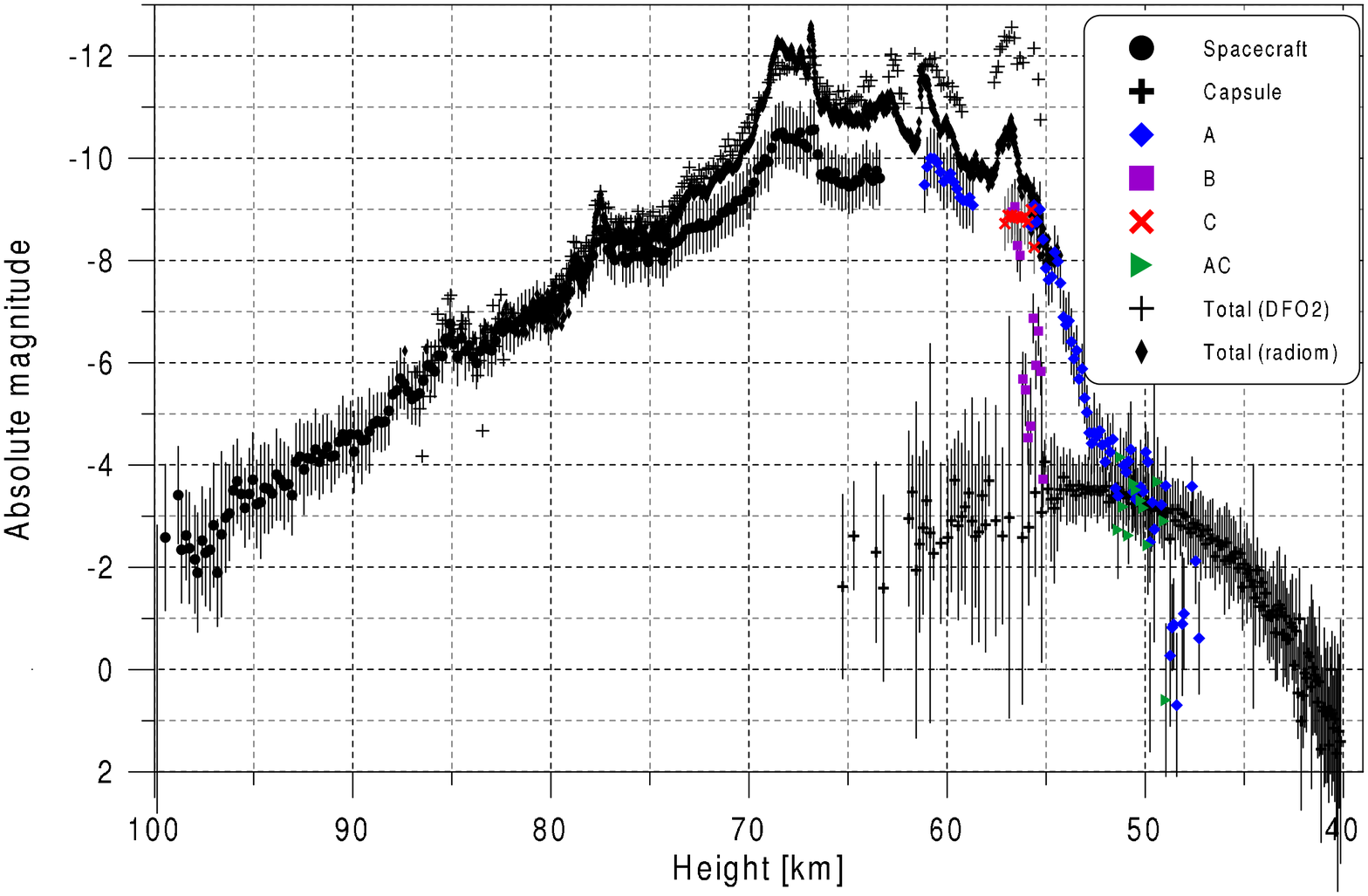}
\caption{Absolute magnitude of the spacecraft, its fragments and the capsule as measured on the GOS3 photograph as a function of height and the total brightness of the fireball as measured on DFO2 photograph and DFO2 radiometer. Formal error bars are given for GOS3. Note that the brightness near maximum is underestimated on the GOS3 image because of saturation and the brightness toward fireball end is overestimated on DFO2 image because of uncertain calibration corrections near the horizon.}
\label{mag}
\end{center}
\end{figure*}

The measured absolute magnitude as a function of height is given in Fig.~\ref{mag}. The spacecraft was first observed at a height of about 100 km, where it was of magnitude $-2$. The brightness then gradually increased, with two minor peaks at 85.1 and 77.5 km, until the broad maximum which started at  68.5 km. The maximal magnitude of $-12.6$ was reached at a height of 66.9 km.  The fireball brightness then fluctuated
but remained high until the final peak at 56.7 km. During this period, the spacecraft disintegrated. After that, the brightness rapidly decreased.

Table~\ref{peaks} lists the positions of the most pronounced peaks on the light curve. Each peak likely corresponds to a fragmentation event. Our photographs do not show geometrically separated fragments above the height of 61.5 km but the major release of small fragments must have occurred at the main flare at 66.9 km. The trajectories (see Fig.~\ref{difh}) suggest that the major fragments A and B separated most likely at the flare at 62.8 km. The brightness of fragments B and C could be first measured
at heights around 57 km, where both of them were of magnitude about $-9$. Fragment B then weakened while fragment C remained bright until its end.

\begin{table*}
\caption{Positions of local maxima on fireball light curve}
\begin{center}
\begin{tabular}{rrrll} \hline
Time & Height & Mag & Velocity & Dyn. pressure \\
s & km & & km/s & kPa \\ \hline
4.1 & 85.1 & $-6.6$ & 12.05  & 1.1 \\
8.1 & 77.5 & $-9.3$ & 12.05  & 3.6 \\
12.9 & 68.5 & $-12.3$ & 11.9  & 13 \\
13.9 & 66.9 & $-12.6$ & 11.85  & 16 \\
16.2 & 62.8 & $-11.3$ & 11.65  & 27 \\
17.1 & 61.3 &$- 11.8$& 11.6   & 32 \\
19.8 & 56.7 & $-10.8$ & 11.1   & 54 \\
\hline
\end{tabular}
\label{peaks}
\end{center}
\end{table*}

The brightness of the capsule was measured on the GOS3 image. The light curve was smooth with a broad maximum of $-3.5$ mag between heights 55--50 km. The measurements at larger heights are sparse and have large error because the proximity of the capsule trail to the bright spacecraft.

From the predicted apparent magnitudes of the capsule at different stations \citep{Fujita}, it can
be inferred that the predicted absolute (100 km distance) brightness of the capsule was $-5.5$ mag
at a height of 50 km. The actual brightness was by 2 magnitudes (factor of 6) lower. 
The expected brightness of the spacecraft was not published.

\section{Discussion}

Our data confirm that the capsule re-entry trajectory predicted by JAXA was correct within few hundreds of meters. 
It is in agreement with the expected accuracy of the prediction, which was estimated to 
500$\times$200 m ($2 \sigma$) at a height of 50 km. The expected error in velocity, 30 m/s
($3 \sigma$), seems to have been underestimated, since the observed deviation was 100 m/s.
The spacecraft followed almost the same trajectory as the capsule, the most significant difference being that the spacecraft was about one kilometer higher at a given longitude. 

To our knowledge, our observation provided the first wide-field, multi-station photographic data for
a re-entering object with good independent telemetry available. Similar observation of the 
Stardust re-entry used different techniques \citep{stardust}. Our work provided an excellent
validation of astrometric and reduction procedures used in fireball networks for determination of 
trajectories and velocities of natural fireballs.

The Hayabusa re-entry can be compared with the entries of natural meteoroids in several aspects: luminous efficiency, dynamic pressure acting at the moment of fragmentation, and apparent ablation coefficient. 

The integral luminous efficiency is defined as the ratio of the total radiated energy to the initial kinetic energy of the object. Using the radiometric light curve (Fig.~\ref{mag}) and taking into account that zero magnitude meteor radiates about 1500 W at all wavelengths  to all directions \citep{SSR}, the total radiative output of the Hayabusa fireball was determined to $3.6\times10^8$ J. Using the mass
of the spacecraft of about 400 kg and the velocity, relative to the atmosphere, of 11.7 km/s, the luminous efficiency was found to be 1.3\%.  
Note that the used conversion from magnitudes to watts is valid for plasma temperature 4500~K. 
The actual temperatures inferred from Hayabusa spectroscopy were in the range 4500 -- 6000 K 
\citep{Abe}. For 6000 K the radiative output and the luminous efficiency would be about 25\% lower.

For natural fireballs, the luminous efficiency is a poorly known parameter, which depends on meteoroid velocity, mass, and structure. The empirical formula of \citet{Brown2002}, based on more energetic events, predicts 4\% for the Hayabusa energy. On the other hand, the formula of \citet{PecCep}, valid for relatively small meteoroids, gives only 0.5\%. \citet{tauKiruna} gave the average luminous efficiency for type I fireballs 5.57\% and for more fragile type II fireballs 1.35\%. We can therefore say that the luminous efficiency of Hayabusa was in the range expected for natural fireballs of similar parameters.

The total radiative output of the capsule was only $2.8\times10^5$ J. The capsule lost only small part from its initial mass of about 20 kg during the re-entry, but it decelerated from 11.7 km/s to just few km/s during the ablation phase, so it deposited about $10^9 $J into the atmosphere. Only about 0.03\% of this energy was radiated out. The low luminous efficiency was due to the special material used for the ablation shield. 
The spectroscopy revealed that the capsule spectrum was dominated at visible wavelengths by
black body radiation of the capsule surface \citep{Abe}. Without ablation products, no large 
radiating plasma envelope, which normally provides most luminosity of fireballs, developed 
around the capsule.

The dynamic pressure $p=\rho v^2$, where $\rho$ is atmospheric density and $v$ is velocity, 
acts on objects moving in the atmosphere. If the dynamic pressure exceeds the 
material strength, fragmentation occurs. Fragmentation is often accompanied by a flare
on the light curve. The last column of Table~\ref{peaks} lists the dynamic pressures at the
flares during the Hayabusa re-entry.  The pressures are quite low and comparable only 
to the most fragile natural meteoroids of cometary origin \citep{Rio}. 
This is not surprising because the spacecraft was not a compact solid body but in fact 
a quite porous object. Nevertheless,
it may be surprising that no piece survived below 47 km.

Finally, the dynamics (deceleration) of fireballs can be fitted using the classical meteor equations \citep{SSR}. The most interesting parameters of the fit are the ablation coefficient, $\sigma$, and the combination $Km^{-1/3} = \Gamma A \delta^{-2/3} m^{-1/3}$, where $\Gamma$ is the drag coefficient and $A$, $\delta$, and $m$ are meteoroid shape coefficient, bulk density, and mass, respectively. The dynamic mass of the meteoroid can therefore be determined if we know or assume $\Gamma$, $A$, and $\delta$.

The typical ablation coefficient of natural fireballs varies from 0.014 s$^2$/km$^2$ for not-so-much fragmenting type I fireballs to about 0.2 s$^2$/km$^2$ for the most fragile type IIIB fireballs \citep{SSR}. We fitted the dynamics of the capsule and spacecraft fragments B and C. For the capsule, we obtained $\sigma = 0.0014 \pm 0.001$ s$^2$/km$^2$, a very low value, which is, nevertheless,  understandable for a specially designed re-entry object. For fragment B, we got $\sigma = 0.07 \pm 0.02$ s$^2$/km$^2$ and for fragment C extremely high $\sigma = 1.1 \pm 0.1$ s$^2$/km$^2$.

The resulting $Km^{-1/3}$ for the capsule can be separated, for example,  into the following reasonable values: $\Gamma A = 0.7$, $\delta = 0.63$ g/cm$^3$ and $m = 21$ kg (almost constant mass). For fragments B and C we have no idea about their shape and density. They may in fact represent clouds of small fragments. Formally, we assumed $\Gamma A = 1.0$ and $\delta = 2$ g/cm$^3$ for both B and C. Then the mass of B was 0.12 kg at the height of 58 km. Such mass, however, cannot explain the observed high luminosity. Either there were more fragments or the density was about 10 times lower and mass 100 times higher. For C, we obtained 12 kg at 58.5 km, which can provide the observed luminosity at that point. However, the computed mass was rapidly decreasing toward the lower heights while the luminosity remained high. This discrepancy can be explained if the object was continuously flattened by the ablation, i.e.\ the shape coefficient was changing.

\vspace{1ex} \noindent
\textit{Acknowledgements.}
We thank JAXA team members who organized and participated in the recovery operations, especially Kazuhisa Fujita (JAXA) who is the chief coordinator of the ground-based observation, Yoshihiro Kakinami (Hokkaido University), Yasuo Shiba (Nippon Meteor Society), Masaharu Suzuki (GOTO INC.), and other members who assisted in observing at GOS3 and GOS4 stations. The work on the Czech side was supported by GA\v CR grants  no. 205/08/0411 and P209/11/1382. SA was supported by the JAXA and the National Science Council of Taiwan (NSC 97-2112-M-008-014-MY3). PAB would like to thank the Science and Technology Facilities Council. This work was supported under grants ST/F003072/1 and ST/H002464/1. We thank the referee, P. Brown, for useful comments, which
helped us to improve the paper.

\end{document}